# Extraction of Schottky barrier height insensitive to temperature via forward current-voltage-temperature measurements

Liu Changshi

Nan Hu College, Jiaxing University, Zhejiang, 314001, P. R. China, E-mail: lcswl@mail.zjxu.edu.cn

Abstract

The thermal stability of most electronic and photo-electronic devices strongly depends on the relationship between Schottky Barrier Height (SBH) and temperature. In this paper, the possible of thermionic current depicted via correct and reliability relationship between forward current and voltage is consequently discussed, the intrinsic SBH insensitive to temperature can be calculated by modification on Richardson-Dushman`s formula suggested in this paper. The results of application on four hetero-junctions prove that the method proposed is credible in this paper, this suggests that the I–V–T method is a feasible alternative to characterize these hetero-junctions.

1. Introduction

Because the SBH controls the current flow to in electronic and photo electronic device based on junctions, obviously, the SBH of each new junctions most be calculated. One typical method to estimate the SBH is voltage-and temperature-current (I-V-T), by employing $\varphi(I_s, T_0) = \frac{kT_0}{q} Ln(\frac{A \cdot R \cdot T_0^2}{I_s})$ the SBH, $\varphi$, of junctions were commonly extracted. The quantity $I_s$ is the saturation current, k is the Boltzmann's constant, q is the electronic charge, A and R are the area of hetero-junction and the effective Richardson's constant of semiconductor in junctions, Hence, one barrier height at fixed temperature , $T_0$, can be calculated by each pair of data in from of $(I_0, T_0)$. Therefore, there was temperature-dependent barrier height in many references [1-5]. As can be seen in references, the values of $\varphi$ increase with increasing temperature [6-10], however, as the temperature increases, the barrier height of GaAs:Cr and Pd/n-GaSb start to decrease [11,12]. In the voltage-depend SBH, the effective SBH is less than the flatland SBH on n-type semiconductor, due to interface-state-derived short-range band bending. The effective p-type SBH does not depend on the applied bias. Our abilities to understand of the deference response to temperature of the SBH at hetero-junctions have advanced little in the past years. Existing SB technology invariably resort to mechanisms not directly linked to the temperature-depend SBH, not because such a connection does not exist, but because this relationship is too complicated to study. SBH measured by internal photoemission for most of these hetero-junctions are very close to unity, insensitive to the measurement temperature [13].

One therefore looks for a technique which is free from the above limitation. In this paper, a simple method based on the modification on Richardson-Dushman's formula and thermionic current is developed for the determination of the barrier height insensitive to temperature of Schottky barrier diodes. The method leads to a reasonable estimation of the thermionic current because one reliability and correct function to calculate forward current by voltage is recommended in this paper. The method is successfully used to determine the barrier height of four Schottky contacts.

2. Methodology and applications

It is tactful to calculate SBH insensitive to temperature through Richardson-Dushman`s formula [14], in according with the Richardson-Dushman`s formula, the intrinsic Schottky barrier height of individual junction can be derived from thermionic current, $I_0$, defined as the zero-electrical field current at different temperatures and the temperature, T.

$$I_0(T) = ART^2 \exp\left(-\frac{q\varphi}{kT}\right) \tag{1}$$

where $\varphi$ is the intrinsic SBH of hetero-junction. According to the principle of nonlinear least-square fitting, parameters A, R and $\varphi$ can be optimized when the curve of T-dependent $I_0$ is the best fitted. Hence, SBH, $\varphi$, can be known based on equation (1).

The key answer to SBH insensitive to temperature is $I_0$. If junction current is described by the thermionic emission theory as [15]

$$I(V,T) = I_0\left[\exp\left(-\frac{qV}{nkT}\right) - 1\right] \tag{2}$$

Hence, the thermionic current or zero-electrical field (V=0) current $I_0$ given by Eq. (2) is always is zero at any temperatures. However, when there is series resistance, $R_s$, in I-V model, it is impossible to reckon $I_0$ via the function in form of $I(V,T) = I_0\left[\exp\left(\frac{qV - IR_s}{nkT}\right) - 1\right]$ because it is not analytically invertible [16].

For experimental current-voltage of hetero-junctions, the growth rate does not steadily decline in the forward direction, but rather increases very quickly. This is shown in the growth curve by an S-shaped, or sigmoidal. Here, the data of current-voltage for hetero-junctions in the forward direction is may be modeled using three-parameter model based sigmoidal as below

[17]

$$I(V) = \frac{I_{max}}{1 + \exp(\alpha(V_c - V))} \quad (V \succ 0) \tag{3}$$

the so-called logisitc model, where $\alpha \succ 0$ and $0 \prec I \prec I_{max}$ in the forward direction. The curve has asymptotes I=I$_{min}$ as V→0 and I=I$_{max}$ as V→∞, which is, of course, never actually attained. This cause few difficulties in practice, because at the voltage at which growth begins to be monitored we have $I \succ 0$. From (3) it is easily seen that the current is at half of the maximum currant (I=I$_{min}$/2), this occurs when V=V$_c$. Again α acts as a scale parameter on V, thus influencing the growth rate.

As examples of potential application we calculate the I-V for four experimental hetero-junctions from different sources. These hetero-junctions are: Au/n-Si [18], Ni/CdSe/p-Si(001) [19], Se/n-GaN [20], and CoSi2/n-Si(100) [21] published in papers. The experimental I-V curves of these hetero-junctions are illustrated in Figs. 1-4, the best parameters to calculate the current by voltage at high level accuary ? were obtained by fitting the experimental I-V with logisitc model (3) and summarized in Table 1. In Figs. 1-4 the theoretical I-V curves of these four hetero-junctions are compared with experimental results, Figs. 1-4 indicate that the calculated results are in good agreement with experimental data. Because logisitc model (3) reproduce the I–V characteristics of those experiments, according to the logisitc model (3), the thermionic current (zero-field) $I_0$ at different temperatures can be estimated from the function (3), the thermionic current is

$$I_0 = I(V,T)\big|_{V=0} = \frac{I_{max}}{1+\exp(\alpha \cdot V_c)} \qquad (4)$$

Fig. 5 shows the experimental $I_0$ vs T plots for the above hetero-junctions. Nevertheless, the results of traditional Richardson's formula (1) fitting on experimental $I_0$ and the temperature, T, is rather unsatisfactory to describe thermionic emission from hetero-junctions, hence, modification of Richardson-Dushman`s formula is carried out, a scaling of $T^\beta$ is suggested which is different from the traditional R-D scaling of $T^2$,

$$I_0(T) = ART^\beta \exp(-\frac{q\varphi}{kT}) \qquad (5)$$

where $\beta$ is an adjustable parameter. In Fig. 5 the modeled thermionic current against temperature is depicted for determining the SBH of the four hetero-junctions. The values of A×R, $\beta$, and $q\varphi/k$ are optimized from the best nonlinear least-square fitting in the above figures and the SBH of the devices insensitive to temperature are calculated using Eq. (5).

3. Conclusions

It is apparent from the above discussions that the proposed method may be used as an accurate technique to determine barrier height of Schottky contacts via forward current-voltage - temperature, the method is useful for evaluating barrier height insensitive to temperature. The proposed I–V–T method begins by exploring the real response of forward current to voltage at temperature, it then proves mathematically that the calculation of the thermionic current is correct and reliability. Our findings suggest that the traditional thermionic emission law governed by the well-known R-D equation is no longer valid, results in our method predicts a scaling of $T^{\beta}$, which is different from the classical R-D scaling of $T^2$, and $\beta$ is variable. It may be mentioned here that the efficient utilizations of both the thermionic current extracted by good explanation on forward current via voltage and modification on Richardson-Dushman's formula are emphasized. The suggested method can be easily applied to either homoginity or inhomoginity hetero-junctions. Clearly, the proposed technique is not complicate and can be applied by simply knowing the D.C. current-voltage characteristics of the device.

characteristics of $CoSi_2$/n-Si(100) Schottky barrier contacts formed by solid state reaction, Solid-State Electronics, 2000, 44, 1807-1818

Table 1 The best parameter to reproduce experimental current-voltage of hetero-junctions at various temperature, T.

| Junctions | T(K) | $I_{max}$ | $V_c$ | α |
|---|---|---|---|---|
| Au/n-Si | 150 | $1.48\times10^{-5}$ | 0.65 | 18.11 |
| | 170 | $1.75\times10^{-5}$ | 0.63 | 17.50 |
| | 190 | $1.47\times10^{-5}$ | 0.58 | 18.25 |
| | 210 | $1.57\times10^{-5}$ | 0.56 | 18.46 |
| | 230 | $1.61\times10^{-5}$ | 0.52 | 18.61 |
| | 250 | $1.43\times10^{-5}$ | 0.46 | 20.36 |
| | 270 | $1.28\times10^{-5}$ | 0.41 | 21.54 |
| | 290 | $1.43\times10^{-5}$ | 0.37 | 22.5 |
| | 310 | $1.75\times10^{-5}$ | 0.32 | 23.03 |
| Ni/CdSe/p-Si | 360 | $1.48\times10^{-3}$ | 1.05 | 4.08 |
| | 340 | $9.2\times10^{-4}$ | 1.04 | 4.12 |
| | 320 | $5.4\times10^{-4}$ | 1.04 | 4.05 |
| | 300 | $3.0\times10^{-3}$ | 1.01 | 4.11 |
| | 280 | $1.7\times10^{-4}$ | 0.95 | 4.40 |
| | 260 | $9.0\times10^{-5}$ | 0.94 | 4.42 |
| | 240 | $5.0\times10^{-5}$ | 0.90 | 4.74 |
| | 220 | $2.0\times10^{-5}$ | 0.99 | 4.44 |
| | 180 | $2.0\times10^{-5}$ | 1.08 | 4.43 |
| | 160 | $1.0\times10^{-5}$ | 1.0 | 4.80 |
| Se/n-GaN | 130 | $2.26\times10^{-5}$ | 1.85 | 4.97 |
| | 160 | $9.16\times10^{-5}$ | 1.93 | 4.23 |
| | 190 | $2.97\times10^{-5}$ | 1.90 | 3.85 |
| | 220 | $5.22\times10^{-4}$ | 1.69 | 3.81 |
| | 250 | $1.23\times10^{-3}$ | 1.76 | 3.41 |
| | 280 | $2.73\times10^{-3}$ | 1.67 | 3.37 |
| | 310 | $4.42\times10^{-3}$ | 1.82 | 2.68 |
| | 340 | $2.34\times10^{-1}$ | 3.41 | 2.16 |
| | 370 | $2.77\times10^{-1}$ | 1.51 | 3.50 |
| CoSi2/n-Si | 293 | 0.01 | 0.28 | 21.29 |
| | 275 | 0.011 | 0.32 | 21.03 |
| | 261 | 0.0074 | 0.31 | 26.32 |
| | 240 | 0.0073 | 0.33 | 29.14 |
| | 220 | 0.0052 | 0.34 | 36.94 |
| | 200 | 0.0033 | 0.35 | 34.50 |
| | 178 | 0.035 | 0.45 | 34.17 |
| | 161 | 0.038 | 0.45 | 33.84 |
| | 140 | 0.026 | 0.46 | 47.18 |
| | 120 | 0.020 | 0.47 | 55.05 |
| | 100 | 0.011 | 0.47 | 81.19 |

Figure caption:

Fig. 1. Forward current as a function of voltage for Au/n-Si junction at different temperature via experiment and theory.

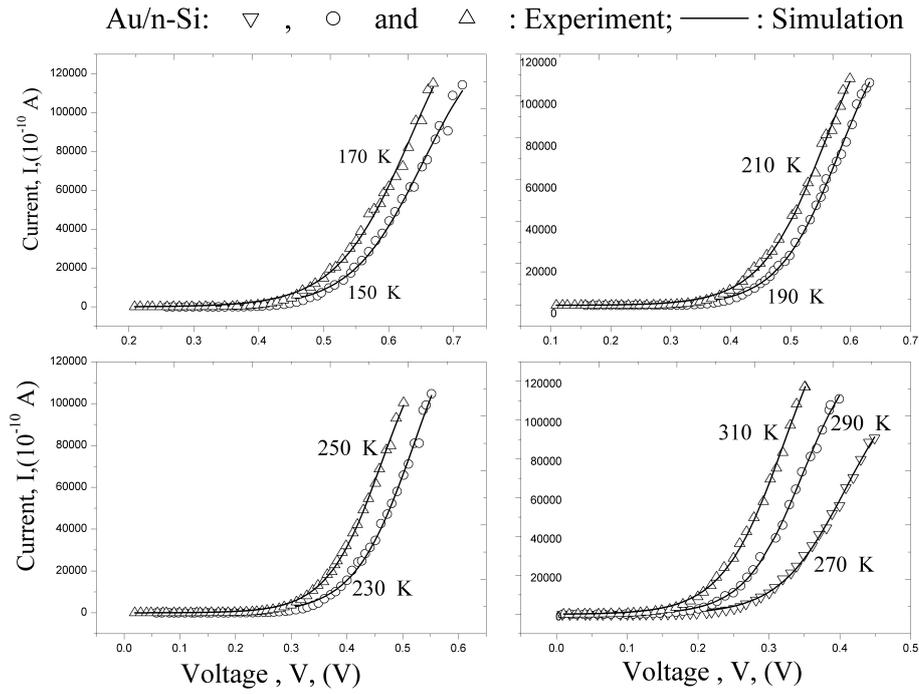

Fig. 2. Forward current as a function of voltage for Ni/CdSe/p-Si(001) junction at different temperature via experiment and theory.

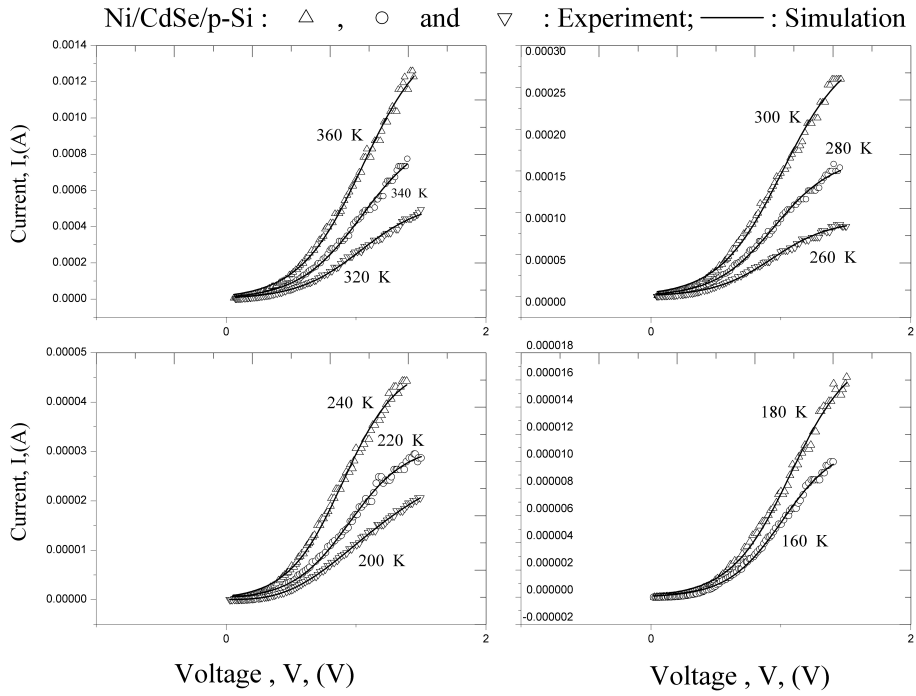

Fig. 3 Forward current as a function of voltage for Se/n-GaN junction at different temperature via experiment and theory.

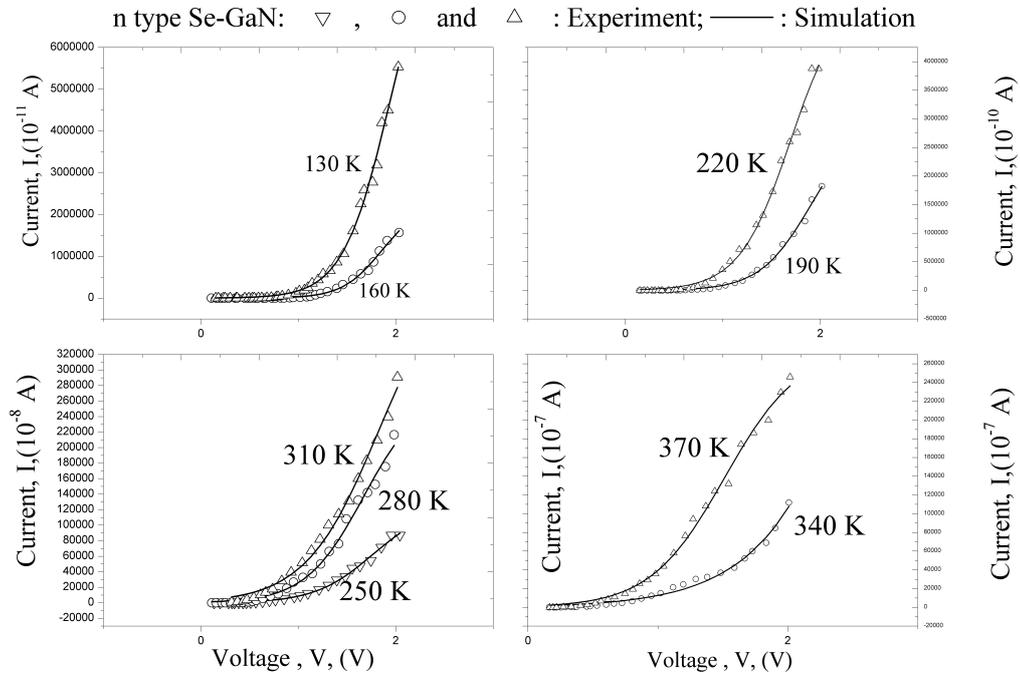

Fig. 4 Forward current as a function of voltage for CoSi2/n-Si(100) junction at different temperature via experiment and theory.

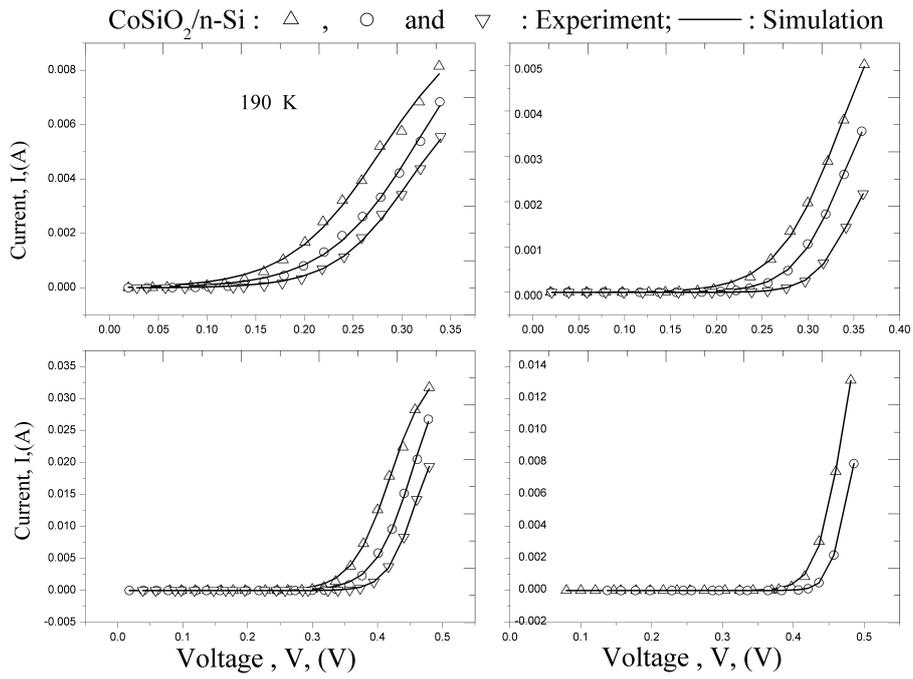

Fig. 5 Thermionic current from Au/n-Si, Ni/CdSe/p-Si(001), Se/n-GaN and CoSi2/n-Si(100) as a function of temperature and fits to the modified Richardson-Dushman`s equation.

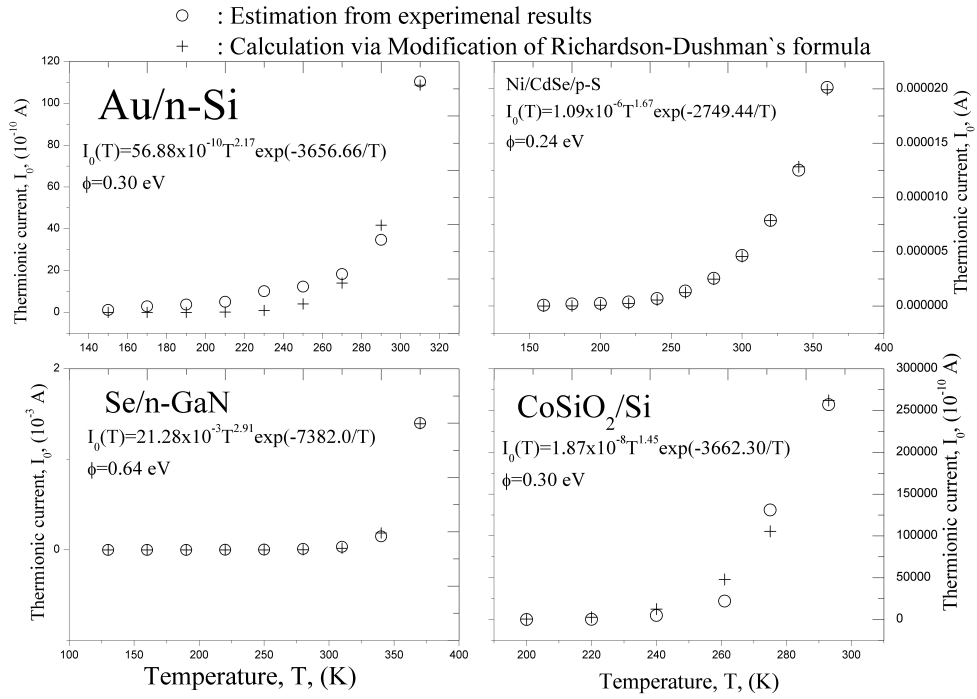